\def\v2  {v$_2$}

\def\um {$\mu\hbox{m}$}

\def\simless{\mathbin{\lower 3pt\hbox
     {$\rlap{\raise 5pt\hbox{$\char'074$}}\mathchar"7218$}}} 
\def\simgreat{\mathbin{\lower 3pt\hbox
     {$\rlap{\raise 5pt\hbox{$\char'076$}}\mathchar"7218$}}} 

\documentstyle[11pt,paspconf]{article}
\def\insertplot#1#2#3#4#5#6#7{
\vskip 10pt\nobreak\hbox to \hsize{\hss\dimen0=#3in\hbox to #6\dimen0{%
\dimen0=#2in\vbox to #6\dimen0{\vss
\special{ps: plotfile #1}
\special{ps::[end]
  PGPLOT restore
}
}\hss}\hss}\vskip 10pt}

\begin{document}

\title{CIRCUMSTELLAR DISKS AROUND PRE-MAIN--SEQUENCE STARS:\\ WHAT ISO CAN
       TELL US}

\author{A. Natta}

\affil{Osservatorio Astrofisico di Arcetri, Largo E. Fermi 5
    I-50125 Firenze, Italy}

\author{ M.R. Meyer and S.V.W. Beckwith}
\affil{ Max Planck Insitut f\"ur Astronomie, Heidelberg, Germany}



\begin{abstract}
ISO observations will improve our understanding of disks around
low-mass pre-main--sequence stars in various ways.
In particular, ISO  can measure
simultaneously
spectral energy distributions  and low-resolution spectra from 2 to
12 \um\
of T Tauri stars
with
greater  sensitivity, greater 
spectral coverage and finer spectral
resolution than available in ground-based observations or with IRAS.
We  illustrate the importance of such observations
for our understanding of the disk 
structure and heating mechanisms,
the disk evolution (in particular the formation of gaps 
caused by  the presence of companion stars or large planets) 
and  dissipation using  preliminary results from PHOT and
PHOT-S for T Tauri stars in the Chamaeleon cloud.

%

\end{abstract}




\section{Introduction}
Circumstellar disks around low mass pre-main--sequence
stars (T Tauri stars, TTS) have been at the center of  attention for
quite a long time.
Disks are believed to account for the infrared and millimeter wavelength
excess emission seen in TTS; if accretion of matter from the disk onto
the star occurs, some of the accretion energy can be  
emitted as hot continuum radiation explaining the
observed  UV excess (see, for example, Hartigan et al. 1995); accretion in a
keplerian disk, coupled to strong magnetic field, can lead to
ejection of matter from the system  and couple the angular momentum of the
star to that of the disk, explaining the puzzling behaviour of the rotation
periods of TTS (see the recent review by Edwards 1997).

The picture of the distribution and evolution of the matter
surrounding low-mass young stars which   emerges from  the wealth of
recent studies is quite complex.
The infrared properties of TTS have been particularly important in shaping
our ideas about disks. When compared to the
observations,  the infrared spectral energy
distribution (SED) of a standard (i.e., geometrically thin, optically thick)
model
disk, either heated by dissipation of viscous energy or by the stellar radiation,
tends to produce too much flux at short wavelengths, and too little at long
wavelengths. The first problem is solved by inserting a hole in
the inner accretion disk (Meyer, Calvet \& Hillenbrand 1997).
Such holes are predicted in 
magnetospheric accreting models where  the stellar magnetic
field disrupts the inner parts of the disk up to distances from
the star of  few stellar radii and channels the accreting matter along
magnetic flux lines (see Kenyon, Yi \& Hartmann 1996
and references therein). The  observed excess emission
at long wavelengths in comparison to model predictions may be  accounted for
(at least partially)
by the flaring of disks at large distances from the star (Kenyon
and Hartmann 1987); other effects of importance are the action
of extended layers of optically thin dust associated with
the optically thick disk,
possibly due to disk winds (Natta 1993) or residual infalling matter
(Natta 1993; Calvet et al. 1994; D'Alessio et al. 1997), and 
the formation of a disk atmosphere in hydrostatic equilibrium (Chiang
and Goldreich 1997; see also Calvet et al. 1992).

In this paper, we will briefly discuss the areas in which ISO will
provide signficant contributions to our understanding of the disk
physical structure,  lifetime and dissipation.

The imaging capability of ISO has been used to 
survey  regions of star  formation
at different wavelengths. These surveys are expected
to yield a complete census (to lower sensitivity limit than ground-
based observations) of young stars {\it with} circumstellar
disks in several regions of star formations.
Nordh et al. (1996; this conference) have presented the first results
of a survey at 6.75 and 15 \um\  of the Chamaeleon region
obtained with CAM
which illustrates  the potential of ISO in this field.
We  refer to their work for further details.

Several ISO programs plan to measure the SEDs of well-known TTS. They plan
to take advantage of the greater ISO sensitivity, greater 
spectral coverage and finer spectral
resolution than available in ground-based observations or with IRAS.
It is in many cases important to determine the shape 
of the SED simultaneously
at all wavelengths, a capability that only ISO offers. 
These programs have the aim of understanding the disk structure and
heating mechanisms,
the disk evolution (with particular interest for the  formation of gaps 
caused by   companion stars or large planets) 
and  dissipation.
As an example of such studies, we present in this talk 
some  preliminary results from
a study of  TTS in Chamaeleon (Beckwith et al, 1998). 

\section {SEDs of T Tauri Stars in Chamaeleon}

The  Chamaeleon sample includes 16 well known TTS.
The positions of the stars on the HR diagram are shown in Fig.~1, 
together with evolutionary tracks
and isochrones from D'Antona and Mazzitelli (1994). 
The 16 stars span approximately
an age
range from $3\times 10^5$ to $10^7$ yrs; 6 are known binaries, and
5 are single (Chelli et al. 1988;
Chen and Graham 1992; Ghez et al. (1997); 
Beckwith et al. 1998)
For all 16 stars we measured broad-band fluxes
in 11 filters, from 4.8 \um\ to 200 \um, with an aperture of 52 arcsec.

\begin{figure}
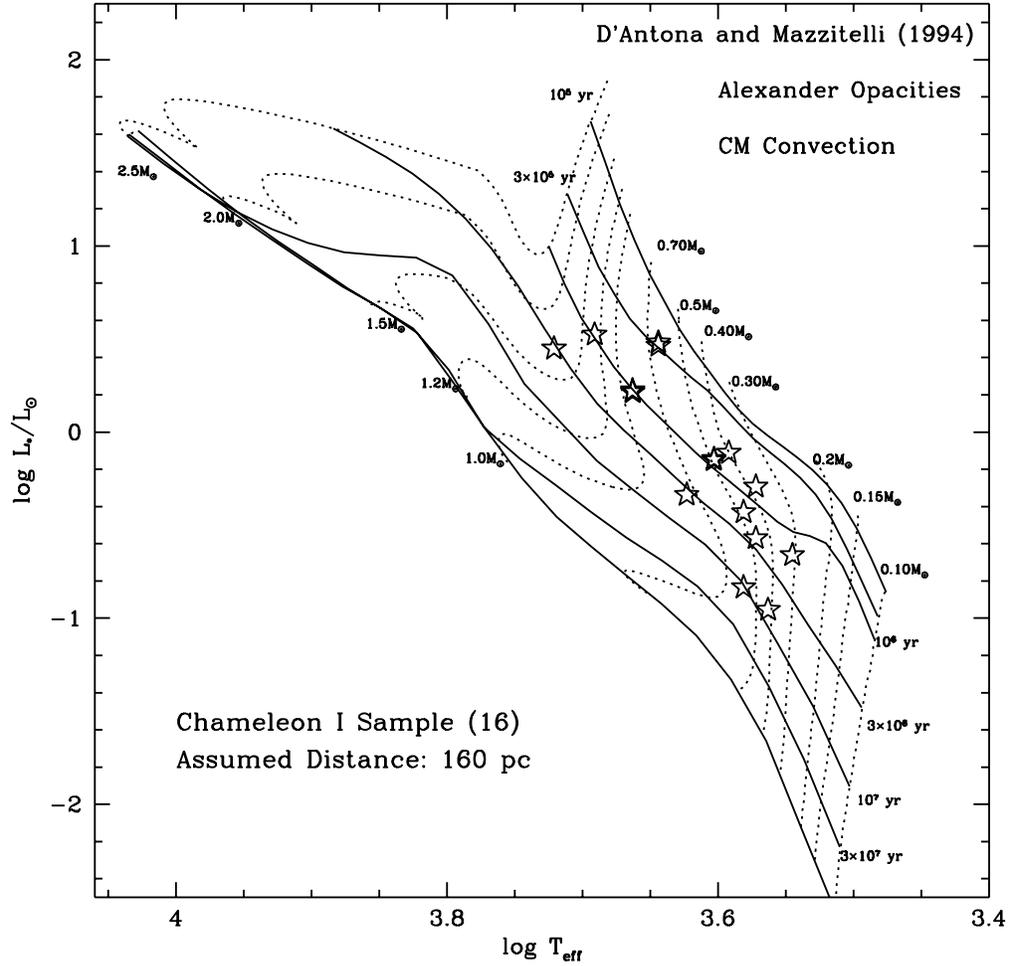

\insertplot{ps.fig1}{8}{10}{-0.5}{2.2}{0.7}{0}
\caption{Location of the  TTS in Chamaeleon
in the ISO sample of Beckwith et al. (1998)
on the HR diagram. Also shown are the D'Antona and
Mazzitelli (1994) evolutionary tracks for stars of different mass
(dotted lines; each track is labelled with the corresponding mass value)
and isochrones (solid lines).}
\end{figure}

Although the calibration uncertainty which affects
these observations  is still large,
some
general features emerge clearly from the data. If
we compare the SED to the predictions of a passively
heated (i.e., with no accretion), geometrically thin disk, 
we find that in all stars 
the emission at short wavelengths is consistent with
the disk model (including, perhaps, the presence of inner holes
for some objects). In contrast,
at long wavelengths the observed fluxes are much larger than
the disk model predicts,
by a factor from  5 to 50, depending on the star.
The transition seems to occur at $\lambda \sim 10$ \um.

The long wavelength excess
is seen in all  16 stars, including those  that 
have very small excess emission at short wavelengths.
The most extreme case in our sample is that of
CS~Cha (see Fig.~2), which shows no evidence of emission
above the photospheric level at $\lambda \simless 10$ \um. Nevertheless,
this star has
a rather strong excess at longer wavelengths (about a factor of 10
at 100 \um\ with respect to the prediction of a passive disk). 
CS~Cha has been searched for companions by Ghez et al. (1997)
in the separation range 0.1-12 arcsec
 and found to be single.
If its extreme SED is interpreted as the emission of a circumstellar disk,
then
the inner disk must have dissipated up
to a distance  of about 0.3 AU from the star. 

\begin{figure}
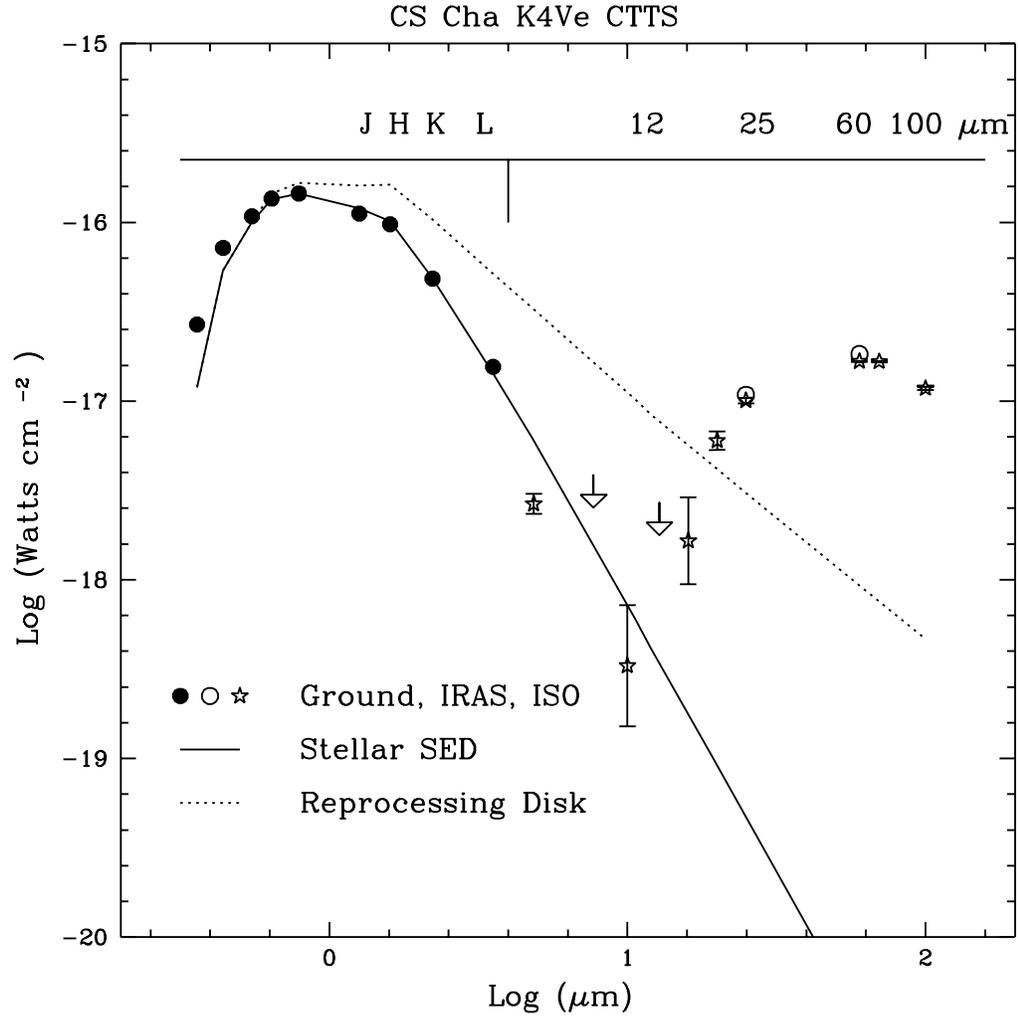

\insertplot{ps.fig2}{8}{10}{-.5}{2.2}{.7}{0}
\caption{ CS~Cha SED. The ISO measurements are shown by stars. The two
arrows denote upper limits from ISO observations. Filled
circles are ground-based photometry. Open circles are IRAS measurements at
25  and 60 \um. The solid curve
shows the SED of the stellar photosphere; the dotted curve the 
model prediction
for a standard reprocessing disk seen face-on.}
\end{figure}

\begin{figure}
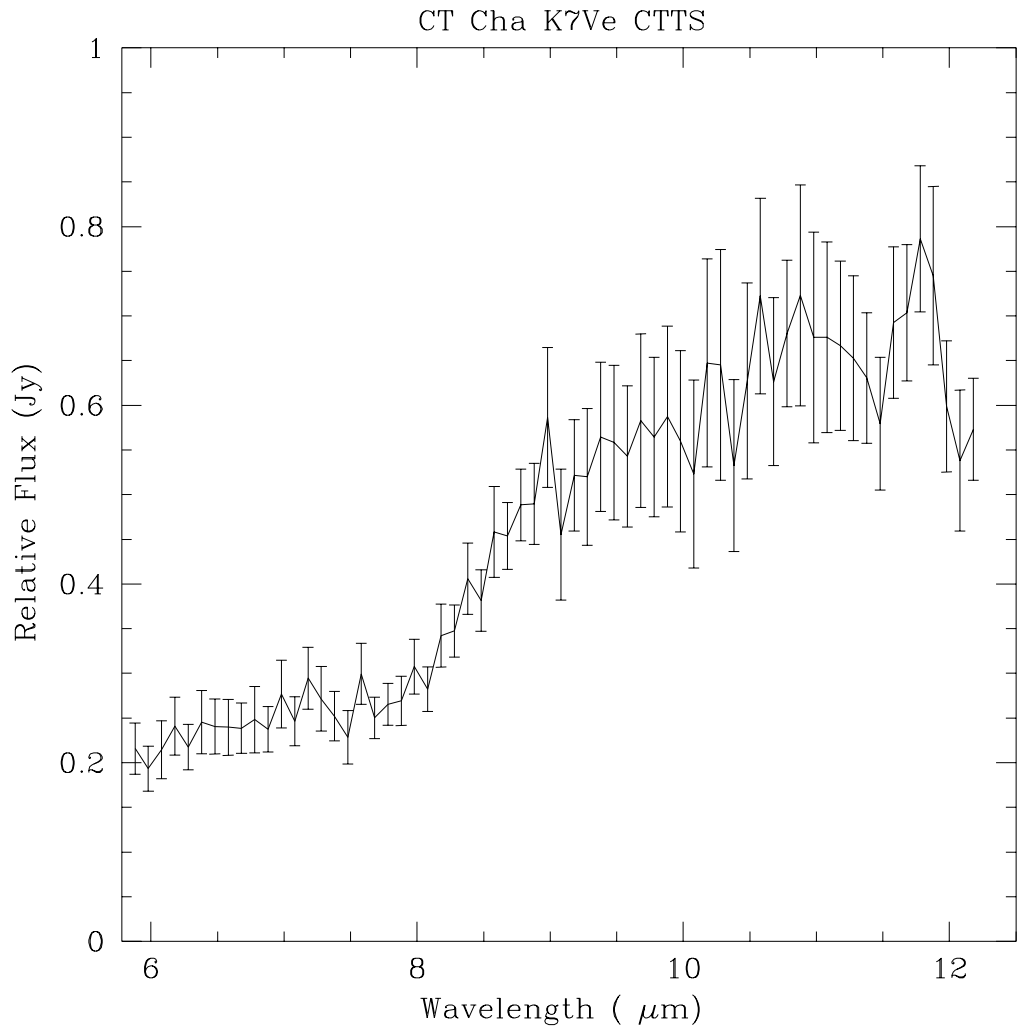

\insertplot{ps.fig3}{8}{10}{-.5}{2.2}{.7}{0}
\caption{ PHOT-S spectrum in the region 6 to 12\um\ of CT~Cha. 
The absolute calibration of the continuum level may be uncertain by
30\% or more.}
\end{figure}

Objects with optically thin inner disks and optically thick remnant
outer disks such as  CS~Cha
are relatively rare (Skrutskie et al. 1990; Wolk \& Walter 1996).
This suggests that the timescale for transition from an inner
accretion disk to a remnant passive disk is very short ($\simless 10^6$ years).

\section {10 \um\ Silicate Feature in TTS}

For 8 of the TTS in Chamaeleon for which we measured the
SED, we  also obtained PHOT-S low-resolution spectra
in the 10 \um\ region. 
In 4 cases out of 8  we  detect the silicate feature
in emission, with values of the peak over continuum ratios as high as 
$\sim$3.
We list the 8 stars in Table 1;
a ``y'' in Column 2 indicates 
that the silicate feature is clearly in emission, a ``n'' that we
see continuum emission only. In no case do we see the silicate
feature in absorption. Table 1 gives also the luminosity and spectral type
of the stars. In the last column, we note the presence or
absence of companions from the references listed above.
Fig.~3 shows  the PHOT-S spectrum between 6 and 12\um\ of the star
CT~Cha.

The presence of the silicate feature in emission in a
a relatively large number of TTS is not new.
In 1985
Cohen and Witteborn (1985) published a survey of 10 \um\ spectra for
pre-main--sequence stars.  19 of
the  27 TTS in their sample showed
the silicate feature in emission, 7 in absorption, 1 a continuum
spectrum.  In several cases, the ratio peak/continuum was
$\sim 2-3$. A similar ratio ($\sim$3) was observed  in the TTS
Elias~28 in Ophiuchus (Hanner et al. 1995).

\begin{table}
\caption{Chamaeleon TTS with PHOT-S Spectra} \label{tbl-1}
\begin{center}
\begin{tabular}{ccccl}
Star & Silicate Em.& Log($L/L_\odot$)& ST& Close Companions \\
\tableline
CT~Cha& y& -0.15& K7Ve& 2.5$"$\\
Glass~I & y& 0.21& K4Ve& 2.5$"$ (IRc)\\
LkH$\alpha$~332-20 & y& 0.52& K2Ve&Sing.\\
SX~Cha& y& -0.29& M0.5Ve& 2.1$"$ (IRc)\\
VZ~Cha& n& -0.34& K6Ve& Sing.\\
WX~Cha& n:& -0.11& K7Ve& 0.8$"$\\
XX~Cha& n& -0.95& M1Ve& ?\\
VW~Cha& n& 0.47 & K5Ve& 0.7$"$/16$"$\\
\end{tabular}
\end{center}
\end{table}

The presence of the silicate feature in emission reveals the existence
in the immediate environment of the TTS  of optically thin, relatively hot dust.
Its intensity (both the absolute value and the
ratio of the peak intensity to the adjacent continuum)
can provide us  with extremely interesting information
on the  properties of the disk, as 
we  illustrate in the following of this section.

Let us assume, for simplicity, that the TTS dust opacity  in the 10\um\
region does not differ much from that in the diffuse interstellar medium.
Then
the ratio of the peak
($\kappa_{sil}$) to continuum ($\kappa_{cont}$) opacity is about 3.5
(Draine and Lee 1984).
 This is the maximum contrast that one can expect to observe, and it is
close to the
values observed in several stars.
The simplest case one can
consider is that of an isothermal
layer of dust at temperature $T$ and optical depth $\tau_\nu$.
At any given frequency $\nu$, the observed intensity is
given by $I_\nu=B_\nu(T)\>(1-e^{-\tau_\nu})$. If $\tau_{sil},\tau_{cont}\gg 1$,
$I_\nu\sim B_\nu(T)$, and the observed spectrum will just be that
of a black-body at $T$. An emission feature can only be
seen when  $\tau_{sil},\tau_{cont}\ll 1$; in this case, however,  
$I_\nu\sim B_\nu(T)\,\tau_\nu \ll B_\nu(T)$, and both continuum and
feature may be too weak to be detected.
The best possibility of detecting a {\it strong} silicate feature in {\it emission}
requires an optical depth at the peak of the silicate feature 
$\tau_{sil}\sim 0.5$ (i.e., A$_V\sim$5). Note, however, that the feature will appear 
slightly weaker,
with respect to the continuum, than in the dust cross section. If
$\kappa_{sil}/\kappa_{cont}=3.5$ 
and $\tau_{sil}\sim 0.5$, then $I_{sil}/I_{cont}\sim 3$.

If the slab of dust is not isothermal,
the emerging spectrum depends not only on the optical depth and temperature
gradient, but also
on the location of the heating source  with respect to the
observer. Fig.~4 shows  the spectra emerging from a slab of dust
of increasing optical depth 
when the observer sees the dust slab from its cool
side (bottom panel) and from its hot side (top panel). In TTS, we can
expect the first situation if the  star+disk system is surrounded by a 
(quasi)spherical, optically thin
 envelope of dust; the second may correspond to cases
where the layer of dust is heated by the stellar radiation
``from above", as in a disk atmosphere. We can see from Fig.~4 that both
cases can produce emission in the silicate feature. However, in the
first case (i.e., when the dust layer is seen from its cold side), the 
ratio of the peak/continuum intensity is  $\sim \kappa_{sil}/\kappa_{cont}$
only for low values of the optical depth (as in the isothermal case),
when both feature and continuum are quite weak.
As $\tau_{sil}$ increases, the colder  foreground dust 
absorbs the emission of the hotter dust in the background
until,   for $\tau_{sil}\simgreat$ 2,
the feature is seen in absorption.
In other words, a ratio $I_{sil}/I_{cont}\sim \kappa_{sil}/\kappa_{cont}$ 
is found only when  the feature and the continuum are both very weak.
On the contrary, when the slab is seen from its bright side, the strength of the
feature increases with $\tau_{sil}$,  and the condition $I_{sil}/I_{cont}\sim
\kappa_{sil}/\kappa_{cont}$ is verified also for large values of $\tau$,
when both feature and continuum are strong.

\begin{figure}
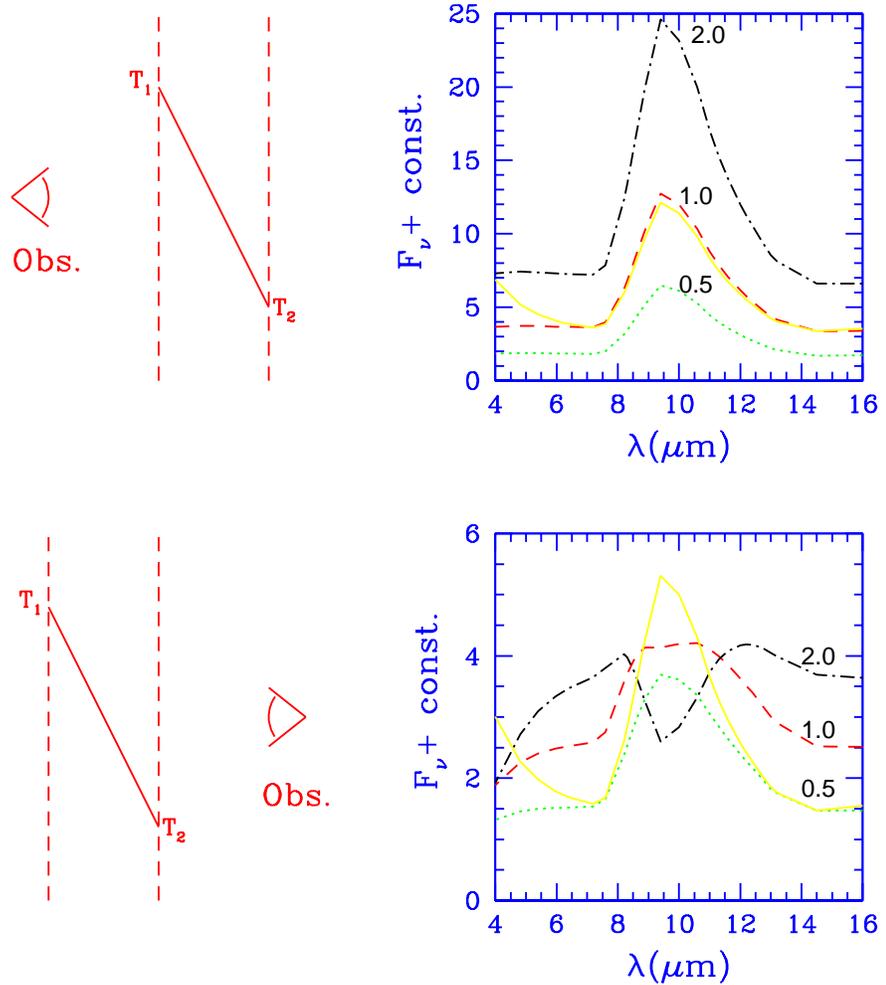

\insertplot{ps.fig4}{8}{10}{-.5}{2.2}{0.7}{0}
\caption{ Emergent intensity of the 10\um\ silicate feature from homogeneous,
plane parallel slabs seen from the hotter side (top panel) and
from the colder side (bottom panel). The temperature decreases
as $R^{-0.5}$
from $T_1$=1000 K to $T_2$=100 K. A sketch of the geometry is
given on the left side of each panel; the resulting spectrum is shown on the
right. In each case, we show the spectrum corresponding to three different
optical depth at the wavelength of the peak of the feature, $\tau_{sil}=0.5,
1$ and 2, as labelled. In both panels the solid curve shows the adopted 
opacity $\kappa_\nu$ as a function of $\lambda$ (arbitrarily normalized).}
\end{figure}

There are few calculations of the expected infrared emission in TTS which
include predictions of the intensity of the 10\um\ silicate feature.
Natta (1993) computes models
where an optically thin spherical envelope of dust surrounds a star+standard
disk system; the envelope spectrum shows the silicate feature in emission,
with a ratio $I_{sil}/I_{cont}\sim$3; however, at
these wavelengths the envelope emission
is
only a  fraction of the emission of the optically thick disk in its
centre, and the
intensity of the feature in the resulting spectrum is only
$\sim$35\% of the continuum. More recently,
Chiang and Goldreich (1997) compute the SED  of
an hydrostatic, radiative equilibrium
model of a  passive disk; in this model,
the disk is surrounded by an optically thin layer of superheated
dust, which, when seen face-on, is reminiscent of
the layer of dust "seen from the hot side" we have shown in Fig.~4,
top panel.
As expected, the predicted ratio $I_{sil}/I_{cont}\sim$3. The continuum
emission in the 10\um\ region is quite strong, and the overall SED
appears  flat over a large range of wavelengths.
A different model is proposed by
Mathieu et al. (1991,1995) for the star GW~Ori. 
This double star shows a very strong
silicate feature which lies in a dip of the continuum, and has a ratio 
peak/continuum of about
2.6. This is well accounted for by a model where a large region of
the disk (from 0.17 to 3.3 AU) is cleared by the dynamical action of
the  companion star; a small amount of dust  (having an 
optical depth at the peak of the silicate feature of
about 0.6)
is left in the region,
and is heated  by the star to $T\sim 560$ K.
In this way, the continuum emission of the optically thick disk is suppressed
and the intensity of the silicate feature in the 
emerging spectrum is maximized.
A typical feature of this model is that the continuum emission 
in the 10 \um\ region is
lower than predicted by  standard, optically thick  disk models,
and the silicate feature lies in a ``valley" of the SED.

ISO observations, which yield simultaneous measurements of the silicate
feature and of the continuum in the 5-25 \um\ region with unprecedented
wavelength coverage, should be capable of discriminating between different
models.

\section {Summary}

ISO observations will contribute to our understanding of disks around
low-mass pre-main--sequence stars in various ways.

Several ISO programs plan to measure the SEDs of well-known TTS. 
We have discussed the importance of such programs
for our understanding of the disk 
heating mechanisms,
the disk evolution,  the  formation of gaps 
caused by  the presence of companion stars or large planets and
the disk dissipation using  preliminary results from PHOT and
PHOT-S observations of TTS in the Chamaeleon cloud.
We have  also discussed
 in some detail the importance of simultaneous observations of the
10 \um\ silicate feature and of the broad-band continuum in the 5-20 \um\
wavelength interval.

%

\end{document}